\documentclass{ws-ijmpa}
\usepackage[super,compress]{cite}
\usepackage{graphicx}
\RequirePackage{color}
\RequirePackage[colorlinks=true
,urlcolor=blue
,anchorcolor=blue
,citecolor=blue
,filecolor=blue
,linkcolor=blue
,menucolor=blue
,pagecolor=blue
,linktocpage=true
,pdfproducer=medialab
,pdfa=true
]{hyperref}

\def\figureautorefname~#1\null{Fig.\,#1\null}
\def\equationautorefname~#1\null{Eq.\,(#1)\null}

\def\m1{M_1}
\def\m2{M_2}
\def\m3{M_3}

\def\ch10{\tilde \chi^0_1}

\newcommand{\lsim}{\mathrel{\mathop{\kern 0pt \rlap
  {\raise.2ex\hbox{$<$}}}
  \lower.9ex\hbox{\kern-.190em $\sim$}}}
\newcommand{\gsim}{\mathrel{\mathop{\kern 0pt \rlap
  {\raise.2ex\hbox{$>$}}}
  \lower.9ex\hbox{\kern-.190em $\sim$}}}
\begin{document}
\markboth{Ning Chen \textit{et~al.}}{New physics implication of Higgs precision measurements}

\catchline{}{}{}{}{}
%
%%%%%%%%%%%%%%%%%%%%%%%%%%%%%%%%%%%%%%%%%%%%%%%%%%%%%%%%%%%%%%%%%%%%

\title{New physics implication of Higgs precision measurements
}

\author{Ning Chen }

\address{School of Physics, Nankai University, Tianjin 300071,China\\
ustc0204.chenning@gmail.com}

\author{Jiayin Gu}

\address{%DESY, Notkestra{\ss}e 85, D-22607 Hamburg, Germany\\
PRISMA Cluster of Excellence, Institut f¨¹{\"u}r Physik, Johannes Gutenberg-Universit{\"a}t,\\
 55099 Mainz, Germany\\
jiagu@uni-mainz.de
}

\author{Tao Han }

\address{PITT PACC, Department of Physics and Astronomy, University of Pittsburgh, \\
 Pittsburgh, PA 15260, USA\\
than@pitt.edu}

\author{Honglei Li }

\address{School of Physics and Technology, University of Jinan,
Jinan, Shandong 250022, China\\
sps$\_$lihl@ujn.edu.cn}

\author{Zhen Liu }

\address{Theoretical Physics Department, Fermi National Accelerator Laboratory, \\Batavia, IL 60510, USA\\
Maryland Center for Fundamental Physics, Department of Physics, University of Maryland, \\College Park, MD 20742, USA\\
zliu2@fnal.gov}

\author{Huayang Song }

\address{Department of Physics, University of Arizona, Tucson, AZ 85721, USA\\
songhuayangfd@gmail.com}

\author{Shufang Su }

\address{Department of Physics, University of Arizona, Tucson, AZ 85721, USA\\
shufang@email.arizona.edu}

\author{Wei Su\footnote{Speaker} }

\address{CAS Key Laboratory of Theoretical Physics, Institute of Theoretical Physics,\\
Chinese Academy of Sciences, Beijing 100190, China\\
ARC Centre of Excellence for Particle Physics at the Terascale, Department of
Physics, \\ University of Adelaide, Adelaide, South Australia 5005, Australia\\
%School of Physics, University of Chinese Academy of Sciences, Beijing 100049, China\\
weisv@itp.ac.cn}

\author{Yongcheng Wu }

\address{Ottawa-Carleton Institute for Physics, Carleton University, 1125 Colonel By Drive, \\ Ottawa, Ontario K1S 5B6, Canada\\
ycwu@physics.carleton.ca}

\author{Jin Min Yang }

\address{CAS Key Laboratory of Theoretical Physics, Institute of Theoretical Physics,\\
Chinese Academy of Sciences, Beijing 100190, China\\
School of Physics, University of Chinese Academy of Sciences, Beijing 100049, China\\
jmyang@itp.ac.cn}

\maketitle

%\begin{history}
%\received{Day Month Year}
%\revised{Day Month Year}
%\end{history}

\begin{abstract}
Studying the properties of the Higgs boson can be an important window to explore the physics beyond the Standard Model (SM). In this work, we present studies on the implications of the Higgs precision measurements at future Higgs Factories.  We perform a global fit to various Higgs search channels to obtain the 95 \% C.L. constraints on the model parameter spaces of Two Higgs Double Model (2HDM) and Minimal Supersymmetric Standard Model (MSSM). In the 2HDM, we analyze tree level effects as well as one-loop contributions from the heavy Higgs bosons. The strong constraints on $\cos(\beta-\alpha)$,   heavy Higgs masses and their mass  splitting are complementary to direct search of the LHC as well as  possible future Z pole precision measurements. For the MSSM, we study both the Higgs couplings and mass precisions. The constraints on the CP-odd Higgs mass $m_A$ and stop mass scale $m_{SUSY}$ can be complementary to the direct search of HL-LHC. We also compare the sensitivity of various future Higgs factories, namely Circular Electron Positron Collider (CEPC), Future Circular Collider (FCC)-ee and International Linear Collider (ILC).

\keywords{Higgs Factories; 2HDM; MSSM}
\end{abstract}

%\preprint{
%\begin{flushright}
%MITP/18-083
%\end{flushright}
%}
%\ccode{PACS numbers:}

%\tableofcontents

\section{Introduction}

While all the indications from the current particle physics measurements seem to confirm the validity of the Standard Model (SM) up to the electroweak scale of a few hundreds GeV, and the observed Higgs boson is SM-like, there are compelling arguments, both from theoretical and observational points of view, in favor of the existence of new physics beyond the SM (BSM). As such, searching for new Higgs bosons would be of high priority since they are present in many extensions of theories beyond the SM.   One of the most straightforward, but well-motivated extensions is the two Higgs doublet model (2HDM) \cite{Branco:2011iw}, as well as Minimal Supersymmetric Standard Model (MSSM). There are five massive spin-zero states in the spectrum ($h, H^0,A^0,H^\pm$) after the electroweak symmetry breaking.
%
%Extensive searches for BSM Higgs bosons have been actively carried out, especially in the LHC experiments%~\cite{Aaboud:2017sjh,CMS:2017epy,Aaboud:2017gsl,Aaboud:2017rel,CMS-PAS-HIG-17-012,Aaboud:2017yyg,Aaboud:2017cxo,ATLAS:2016qiq, ATLAS:2016grc,ATL-PHYS-PUB-2013-016, CMS:2013dga,Baglio:2015wcg}
%. Unfortunately, no signal observation has been reported thus far. This would imply either the non-SM Higgs bosons are much heavier and essentially decoupled from the SM, or their interactions are accidentally aligned with the SM configuration \cite{Carena:2013ooa}. In either situation, it would be challenging to directly observe those states in experiments.

Complementary to the direct searches,  precision measurements of SM parameters and the Higgs properties could lead to relevant insights on new physics.
High precision achieved at future Higgs factories with about $10^6$ Higgs bosons, and possible $Z$ pole measurements with $10^{10} - 10^{12}\ Z$ bosons~\cite{CEPC-SPPCStudyGroup:2015csa,Gomez-Ceballos:2013zzn,Baer:2013cma,ALEPH:2005ab} would hopefully shed light on the new physics associated with the electroweak sector. To take advantage of these precisions~\cite{Gu:2017ckc,Chen:2018shg}, we make a global fit to explore their abilities of detecting new particles and constraining model parameter space.

\section{Study methods and Global fit results}

\begin{figure}[h]
\begin{center}
\includegraphics[width=5cm]{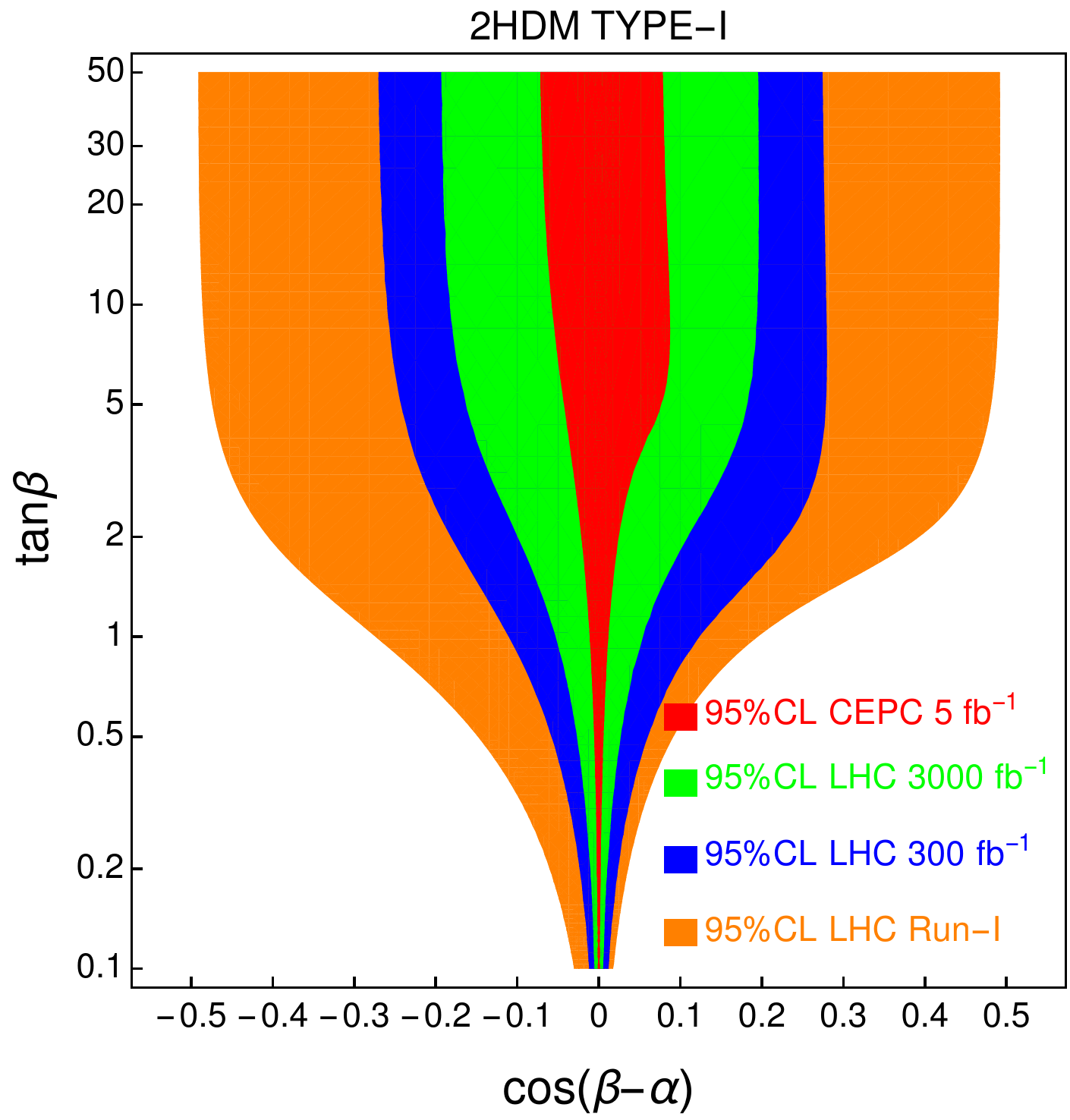}
\includegraphics[width=5cm]{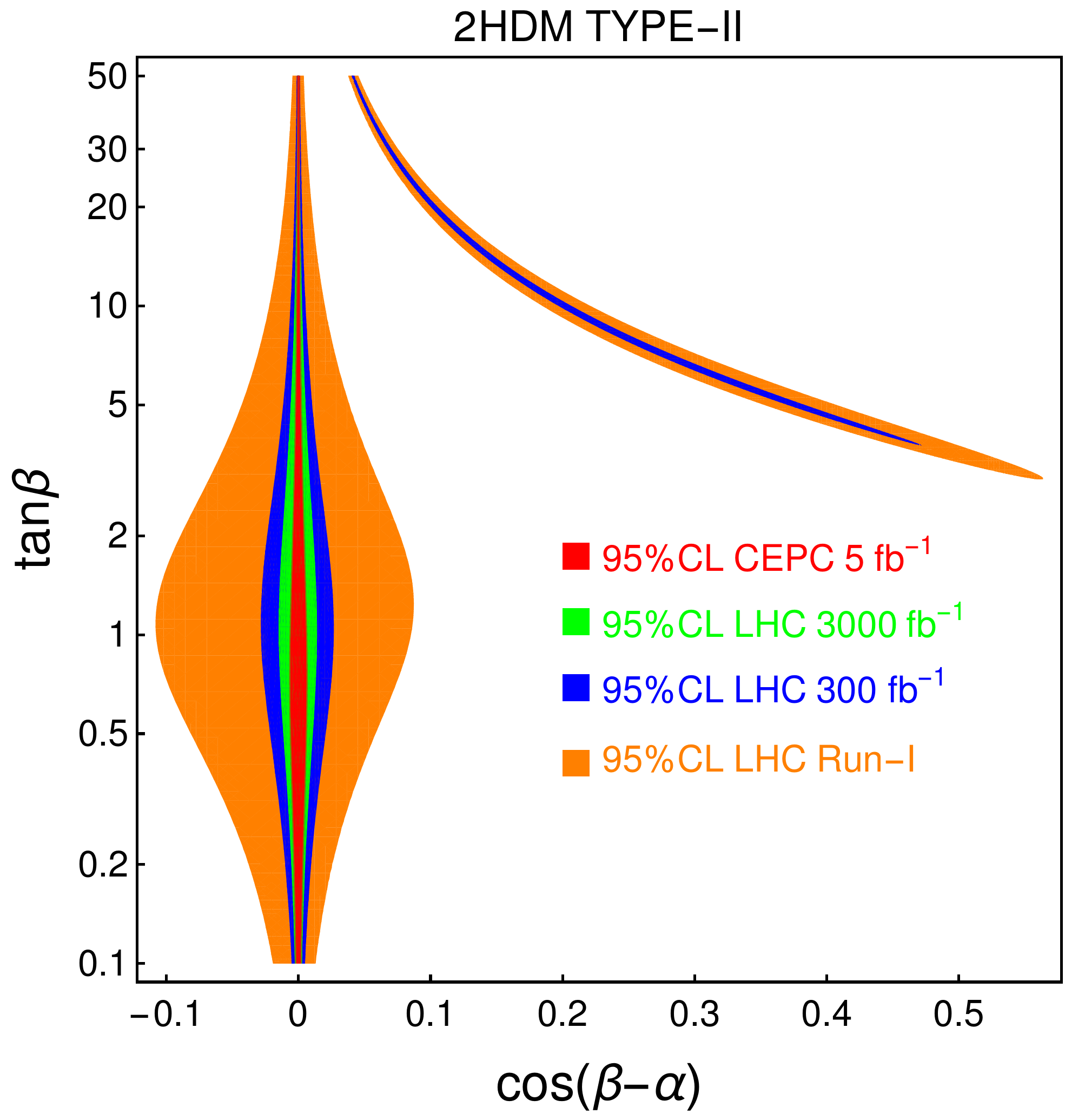}%\\\vspace{3mm}
\caption{The allowed region in the  plane of $\cos(\beta-\alpha)$-$\tan \beta$  at 95\% C.L. for the Type-I and Type-II of 2HDM, given LHC and CEPC Higgs precision measurements.  The other two types are similar to Type-II.  The special  ``arm" regions for the Type-II is the wrong-sign Yukawa region. More details are shown in Ref.~\cite{Gu:2017ckc}.}
\label{fig:cepc-tree}
\end{center}
\end{figure}

There is a plethora of articles in the literature to study the effects of the heavy Higgs states on the Higgs couplings in Models with extended Higgs sector~\cite{Branco:2011iw,Gu:2017ckc,Chen:2018shg}.  In 2HDM,
identifying the light CP-even Higgs $h$ to be the experimentally observed 125 GeV Higgs, the couplings of $h$ to the SM fermions and gauge bosons receive two contributions:  tree-level values, which are controlled by the mixing angles $\alpha$ of the CP-even Higgses  and $\tan\beta$, ratios of the vacuum expectation values of two Higgses: $\tan\beta=v_2/v_1$, and loop contributions with heavy Higgses running in the loop.  % Of particular interest is  the so-called ``alignment limit"~\cite{Carena:2013ooa} of $\cos(\beta-\alpha)=0$, in which the light CP-even Higgs couplings are identical to the SM ones at the tree-level, regardless the other scalar masses.  Loop corrections, however, could lead to deviations of the couplings of $h$ to SM particles, even at the alignment limit.

With a global fit to the Higgs rate measurements at the LHC as well as the CEPC, assuming that no deviation to the SM values is observed at future measurements,   the 95\% C.L. region in the $\cos(\beta-\alpha)$ vs. $\tan\beta$  plane for various types of 2HDM (depending on how the two Higgs doublets are coupled to the quark and lepton sectors)  are shown in~\autoref{fig:cepc-tree} for tree-level only effects. $\cos(\beta-\alpha)$ in all four types are tightly constrained at both small and large values of $\tan\beta$, except for Type-I, in which constraints are relaxed at  large $\tan\beta$ due to suppressed Yukawa couplings.

 \begin{figure}[h!]
 \begin{center}
\includegraphics[width=5 cm]{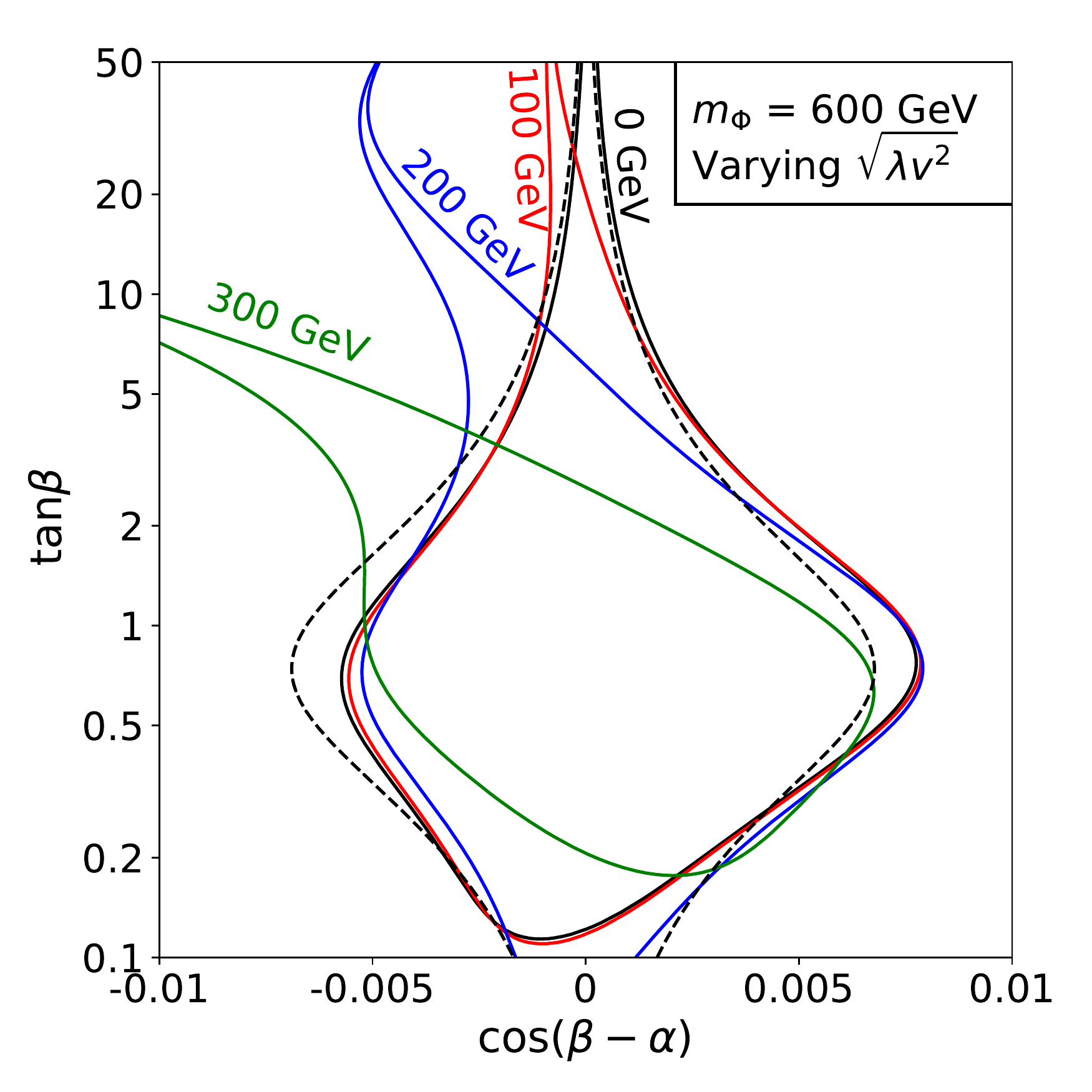}
\includegraphics[width=5 cm]{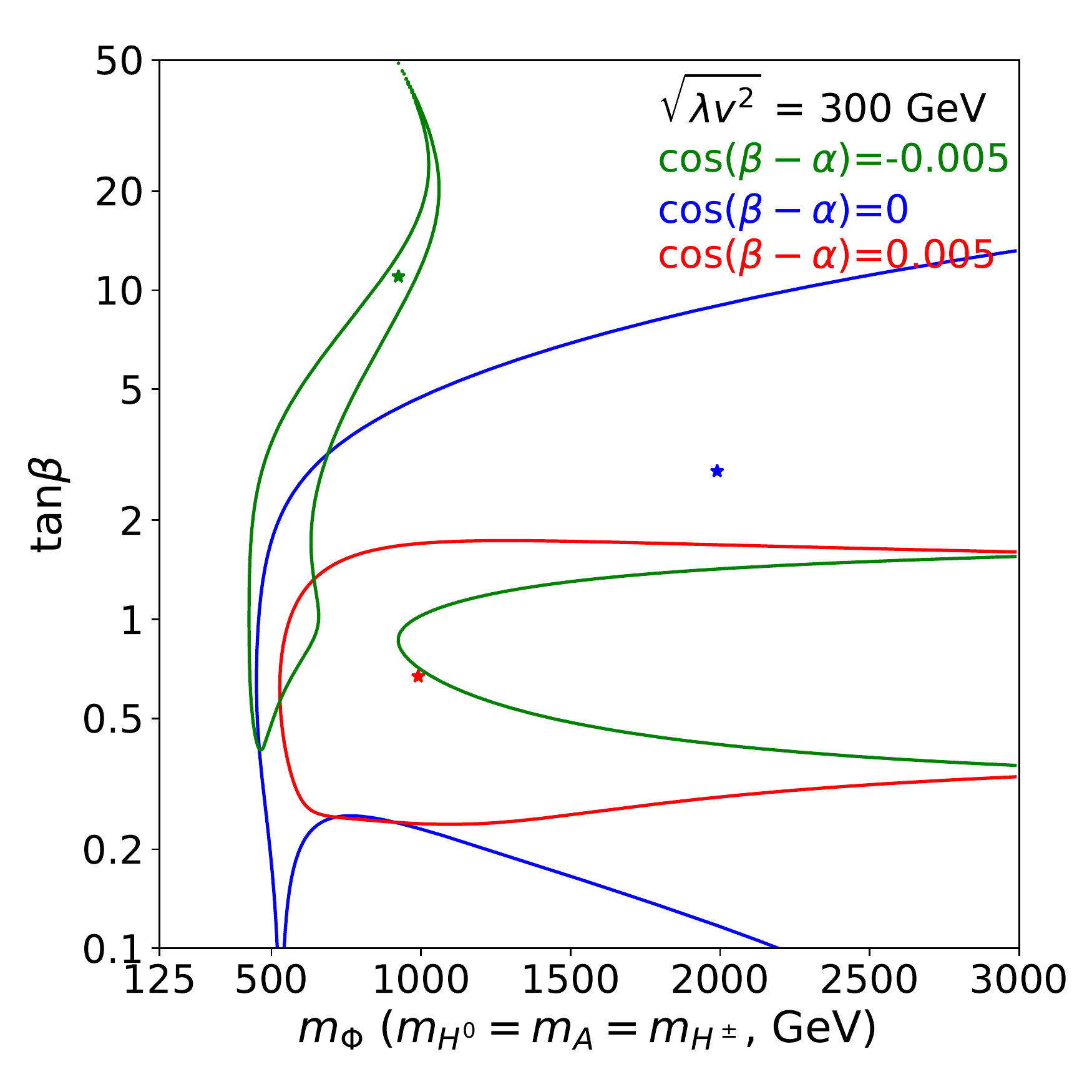}%\\
\caption{Three-parameter fitting results at 95\% C.L. with CEPC precision for Type-II 2HDM.   The left panel shows the parameter space $\cos(\beta-\alpha)$ vs. $\tan \beta$, varying the value $\sqrt{\lambda v^2}$ with $m_A=m_H=m_{H^\pm}=m_\Phi=600$  GeV.   The tree-level only global  fit results are shown by the dashed black lines for comparison.
The right panel shows the $m_\Phi$ vs. $\tan \beta$  plane,  varying the value of $\cos(\beta-\alpha)$ with $\sqrt{\lambda v^2}=300$ GeV.     The colored stars show the corresponding best fit point.  More details are shown in Ref.~\cite{Chen:2018shg}. }
\label{fig:2HDM_cepc}
\end{center}
\end{figure}

To fully explore the Higgs factory potential in search for new physics beyond the SM,  both the tree-level deviation and loop corrections need to be considered.~\autoref{fig:2HDM_cepc} shows the 95\% C.L. global fit results to all CEPC Higgs rate measurements in  the Type-II 2HDM parameter space,  including both tree level and loop corrections.   Degenerate Heavy Higgs masses $m_A=m_H=m_{H^\pm}=m_\Phi$ are assumed such that $Z$-pole precision measurements are automatically satisfied.  The left panel shows $\cos(\beta-\alpha)$ - $\tan\beta$ parameter space with regions enclosed by curves are allowed if no deviation from the SM prediction is observed.  Black, red, blue and green curves are for model parameter $\sqrt{\lambda v^2}=\sqrt{m_\Phi^2-m_{12}^2/s_\beta c_\beta}=0$, 100, 200, and 300 GeV, respectively.   The tree-level only global  fit results are shown by the dashed black lines for comparison.   $|\cos(\beta-\alpha)|$ is typically constrained to be less than about 0.008 for $\tan\beta\sim 1$.  For smaller and larger values of $\tan\beta$, the allowed range of $\cos(\beta-\alpha)$ is greatly reduced.   Loop effects from heavy Higgses tilt the value of $\cos(\beta-\alpha)$ towards negative, especially in the large $\tan\beta$ region.

The right panel of~\autoref{fig:2HDM_cepc} shows the 95\% C.L. allowed region in  $m_\Phi$ - $\tan\beta$ plane, with $\cos(\beta-\alpha)=-0.005$ (green), 0 (blue) and 0.005 (red).  In the alignment limit of  $\cos(\beta-\alpha)=0$,  the heavy Higgs mass $m_\Phi>500$ GeV are still allowed for $\tan\beta \lesssim 10$.   Once deviating away from the alignment limit, the constraints on the heavy Higgs mass get tighter.    Comparing to the direct searches of the heavy Higgs bosons at hadron colliders~\cite{Aaboud:2017sjh},   The reach in the heavy Higgs mass and couplings at future Higgs factories can be complementary to the direct search limits at the LHC, especially for intermediate values of $\tan\beta$.
\begin{figure}[tb]
\begin{center}
 \includegraphics[width=10.cm]{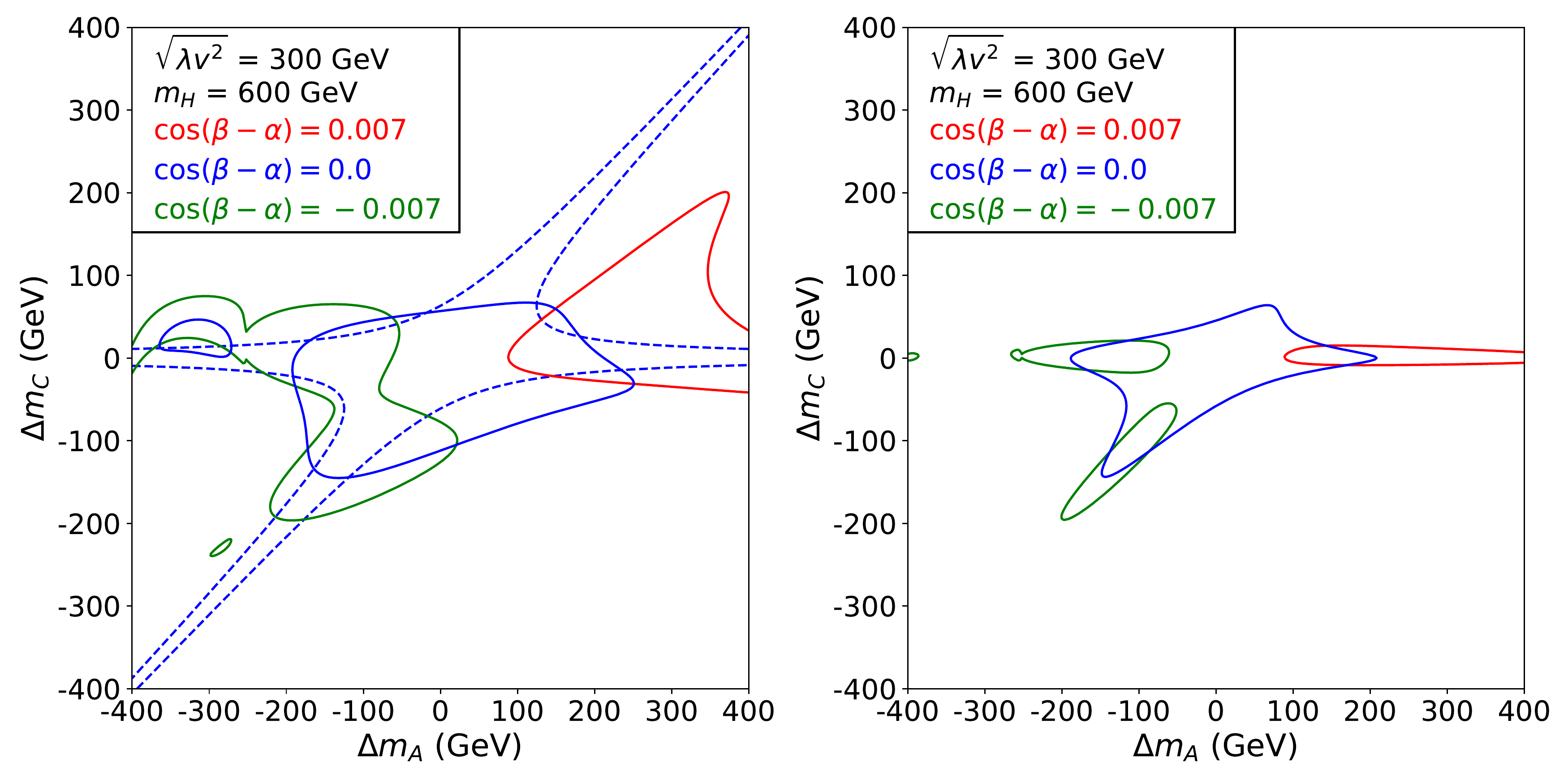}
  \caption{Three-parameter fitting  95\% C.L. range of $\Delta m_A$ - $\Delta m_C$ plane,  focusing on the $\cos(\beta-\alpha)$ dependence (given by different colored lines), for Higgs and $Z$-pole precision constraints individually (left panel),   and combined constraints (right panel) in the Type-II 2HDM.   More details are shown in Ref.~\cite{Chen:2018shg}.   }
\label{fig:da-dc-cos}
\end{center}
\end{figure}

Going beyond the degenerate mass case, both the Higgs and $Z$-pole precision variables are sensitive to the mass splittings between the charged Higgs and the neutral ones.
~\autoref{fig:da-dc-cos} shows the 95\% C.L. range of   $\Delta m_A=m_A-m_H$ vs. $\Delta m_C=m_{H^\pm}-m_H$  plane, for Higgs and $Z$-pole precision constraints individually in (left panel),  and combined constraints (right panel), with  $m_H=600$ GeV  and $\sqrt{\lambda v^2}= 300$ GeV.   For the Higgs precision fit, the alignment limit (blue curve) leads to both $\Delta m_A$ and $\Delta m_C$ around 0 within a few hundred GeV range.   Even for small deviation away from the alignment limit,  $\Delta m_A$ is constrained to be positive for $\cos(\beta-\alpha)= 0.007$, and negative for  $\cos(\beta-\alpha)= -0.007$.   The $Z$ pole precision measurements (shown in region enclosed by blue dashed curves) constrain either $\Delta m_C\sim 0$ or $\Delta m_C \sim \Delta m_A$, equivalent to  $m_{H^\pm} \sim m_{H,A}$.  %The dependence on $\cos(\beta-\alpha)$ is almost non-noticeable given the small range of $\cos(\beta-\alpha)$ allowed under the current LHC Higgs precision measurements.
Combining both the Higgs and $Z$ pole precisions (right panel), the range of $\Delta m_{A,C}$ are further constrained to a narrower range.   The expected accuracies at the $Z$-pole and at a Higgs factory are quite complementary in constraining heavy Higgs mass splittings.
\begin{figure}[h]
\begin{center}
\includegraphics[width=6.cm]{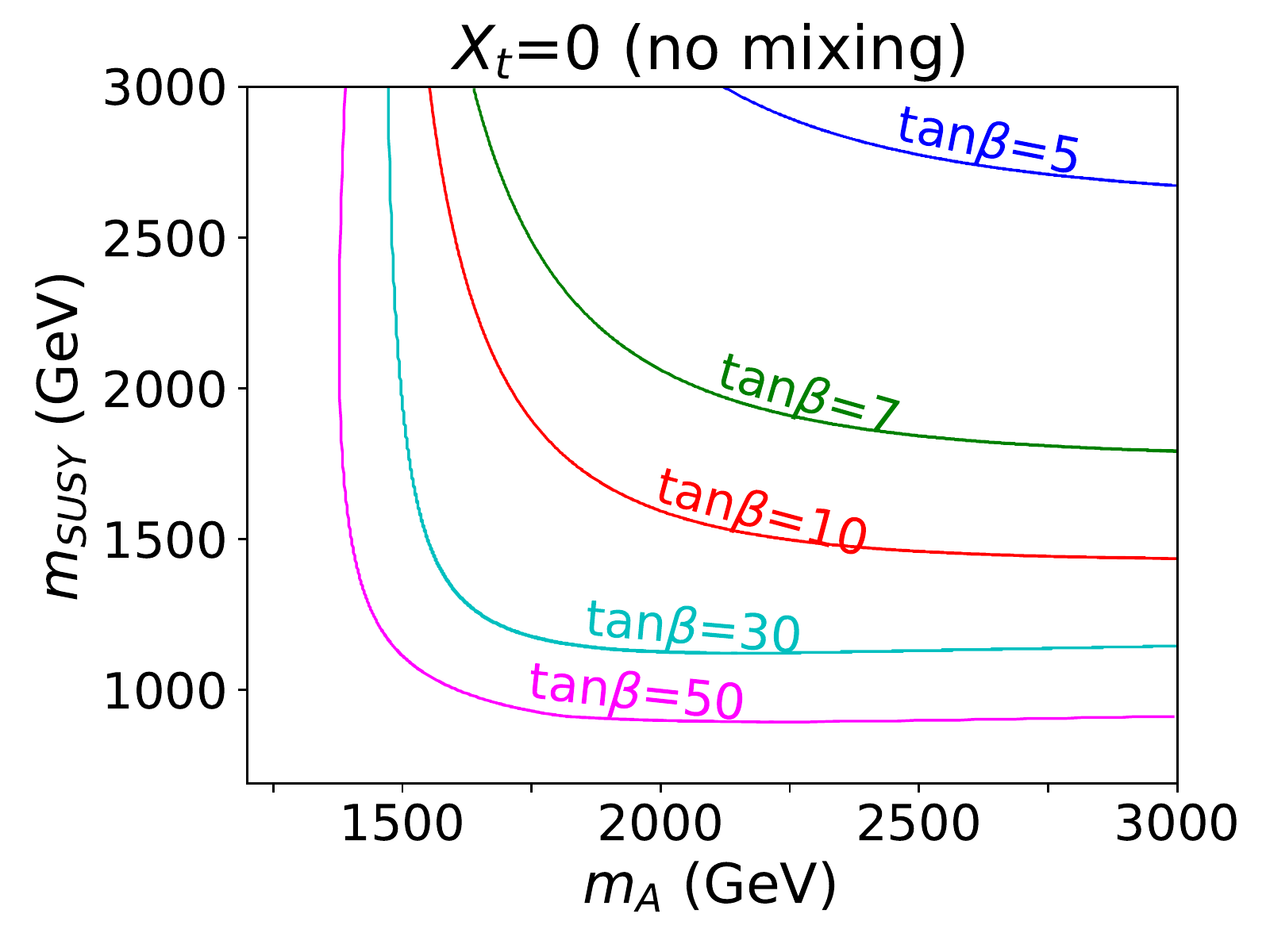}
\includegraphics[width=6.cm]{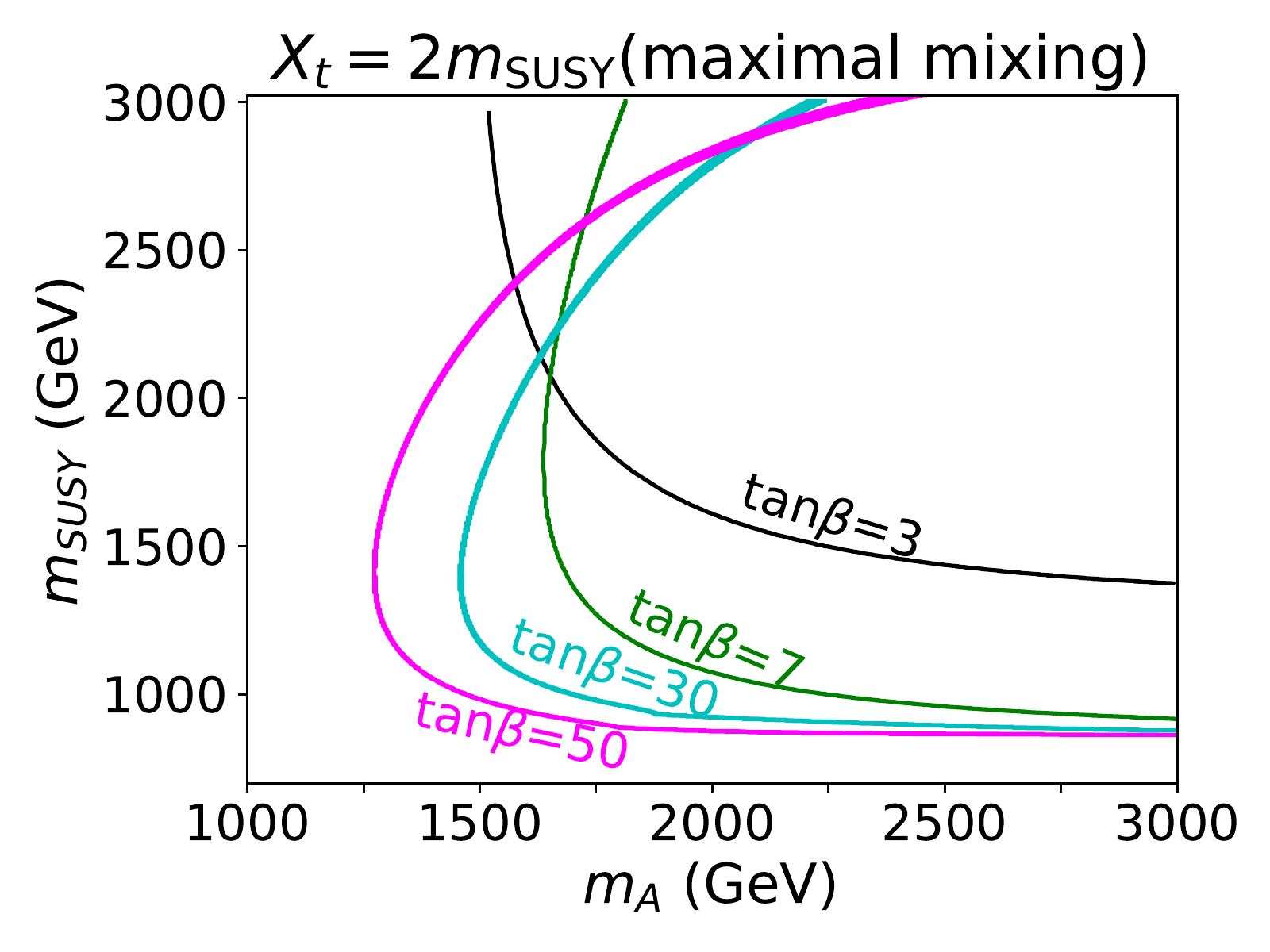}%\\
\caption{The three dimension global fitting with varying $m_A, m_{SUSY}$ and $\tan \beta$ for setting $X_t =0$ ( the left one) and $X_t = 2 m_{SUSY}$ ( the right one) separately.  $\mu$ is set to be 500 GeV.
%The three dimension global fitting with varying $M_A, M_{SUSY}$ and $\tan \beta$ for setting $X_t =0$ (no mixing of stop sector, the left one) and $X_t = 2 M_{SUSY}$ (maximal mixing of stop sector, the right one) separately. Our parameters space are $\mu = 500 \text(GeV), M_A \in (200, 3000) \text{~GeV},  M_{SUSY} \in (200, 3000) \text{~GeV}$ and $\tan\beta \in (1, 50)$. For projection to $M_A .vs. M_{SUSY}$ plane in the upper two panels, we choose serval specific $\tan\beta$ values, 50 for the magenta, 30,10,7,5,3 for the cyan, red, green, blue and black respectively. The lower two are the projection of three dimension global fitting in the plane $M_A$ vs. $\tan \beta$. We choose serval specific $M_{SUSY}$ values. For the $X_t =0$ the red, blue, green, cyan and black are for $M_{SUSY}$ = 2000, 1500, 1000, 950, 850 GeV respectively and for the $X_t =\sqrt 6 * M_{SUSY}$, the red, blue, green, orange and black are for $M_{SUSY}$ = 2000, 1500, 1300, 1200, 1100 GeV respectively. As stated before, there is a special corner at 'large $\tan \beta$ + $M_A \le 200 $ GeV' for the maximal mixing stop sector.
}
\label{fig:mssm1}
\end{center}
\end{figure}

As a specific model of Type-II 2HDM, MSSM   can also be strongly constrained by future Higgs precisions. As in~\autoref{fig:mssm1}, we show the global fit result in plane $m_A-m_{SUSY}$. Here we choose serval $\tan\beta$ values to show results, 50, 30,10,7,5,3 for the magenta, cyan, red, green, blue and black respectively. The left(right) is for no(maximal) left-right stop mixing sector, and we can see $\tan \beta $ less than 5(3) is totally excluded by CEPC Higgs precisions. Generally we can have $m_A \geq 1.2$ TeV, $m_{SUSY} \geq 800$ GeV, which can be complementary to future HL-LHC.

\section{Conclusion}
In this paper, we %studied the impacts
presented the results for the impactsof the precision measurements of the SM parameters at the proposed $Z$-factories and Higgs factories on the extended Higgs sector, including 2HDM and MSSM. For the tree-level 2HDM, $|\cos(\beta-\alpha)|$ can be restricted 0.008. When including the loop effects, CEPC precision can give lower bound on non-SM Higgs masses, as well as their splitting.
Combining the Higgs and Z-pole precisions, the typical heavy Higgs mass splitting is constrained to be less than about 200 GeV. For the MSSM, the constraint on mA is 1.2 TeV, which is complementary to that from the direct searches at future hadron colliders.
%While combing the Higgs and $Z$-pole precisions, the typical heavy Higgs mass splitting is less than about 200 GeV. For the MSSM, the constrain on $m_A$ is 1.2 TeV, which is complementary to future hadron colliders.
\section*{Acknowledgments}
The work is supported by the National Natural Science Foundation of China under Grant No. 11575176, Grant No. 11635009, Grant No.~11675242, by the U.S.~Department of Energy under Grant No.~DE-FG02-95ER40896, Grant No.~DE-FG02-13ER41976/DE-SC0009913/DE-AC02-07CH11359, by the Natural Sciences and Engineering Research Council of Canada (NSERC)

\end{document}